\documentclass[12pt]{article}
\usepackage{amssymb}

\usepackage{amsmath}
\usepackage[unicode,bookmarks,bookmarksopen,bookmarksopenlevel=2,colorlinks,linkcolor=blue,citecolor=green]{hyperref}

\input{tcilatex}

\begin{document}

\title{Integrability of the Egorov hydrodynamic type systems}
\author{Maxim V. Pavlov \\
Lebedev Physical Institute, Moscow}
\date{}
\maketitle

\begin{abstract}
Integrability criterion for the Egorov hydrodynamic type systems is
presented. The general solution by the generalized hodograph method is
found. Examples are given. A description of three orthogonal curvilinear
coordinate nets is discussed from the viewpoint of reciprocal
transformations.
\end{abstract}

\vspace{1cm}

\textit{In honour of Sergey Tsarev}

\vspace{1cm}

\tableofcontents

\vspace{1cm}

\textit{keywords}: Hamiltonian structure, reciprocal transformation, Egorov
metric, hydrodynamic type system, Riemann invariant, extended hodograph
method, generalized hodograph method.

MSC: 35L40, 35L65, 37K10;\qquad PACS: 02.30.J, 11.10.E.

\section{Introduction}

The theory of integrable hydrodynamic type systems%
\begin{equation}
u_{t}^{i}=\overset{N}{\underset{j=1}{\sum }}v_{j}^{i}(\mathbf{u})u_{x}^{j}%
\text{, \ \ \ \ \ }i=1,2,...,N  \label{2}
\end{equation}%
was created by B.A. Dubrovin and S.P. Novikov (see \textbf{\cite{Dubr+Nov}})
and developed by S.P. Tsarev (see \textbf{\cite{Tsar}}).

An integrability of the hydrodynamic type systems (\textbf{\ref{2}}) was
investigated by S.P. Tsarev in \textbf{\cite{Tsar}} for the distinct
characteristic velocities $v^{i}$, which are determined by the algebraic
system%
\begin{equation*}
\det \left| v_{k}^{i}(\mathbf{u})-v\delta _{k}^{i}\right| =0.
\end{equation*}%
If (\textbf{\ref{2}}) can be written via the Riemann invariants $r^{i}(%
\mathbf{u})$ ($i=1,2,...,N$) in the diagonal form%
\begin{equation*}
r_{t}^{i}=v^{i}(\mathbf{r})r_{x}^{i},
\end{equation*}%
then the integrability condition (so-called the ``semi-Hamiltonian''
property) is given by%
\begin{equation*}
\partial _{j}\frac{\partial _{k}v^{i}}{v^{k}-v^{i}}=\partial _{k}\frac{%
\partial _{j}v^{i}}{v^{j}-v^{i}}\text{, \ \ \ \ }i\neq j\neq k,
\end{equation*}%
where $\partial _{k}\equiv \partial /\partial r^{k}$. Any semi-Hamiltonian
hydrodynamic type system (\textbf{\ref{2}}) possesses an infinite set of
conservation laws%
\begin{equation}
\partial _{t}h(\mathbf{u})=\partial _{x}p(\mathbf{u})  \label{cons}
\end{equation}%
and an infinite set of commuting flows (i.e. the functions $u^{k}$
simultaneously are functions of $x,t$ and $\tau $, then the compatibility
conditions $(u_{\tau }^{i})_{t}=(u_{t}^{i})_{\tau }$ must be fulfilled)%
\begin{equation}
u_{\tau }^{i}=\overset{N}{\underset{j=1}{\sum }}w_{j}^{i}(\mathbf{u}%
)u_{x}^{j}\text{, \ \ \ \ \ }i=1,2,...,N  \label{com}
\end{equation}%
parameterized by $N$ arbitrary functions of a single variable (see \textbf{%
\cite{Tsar}}). The algebraic system%
\begin{equation}
x\delta _{k}^{i}+tv_{k}^{i}(\mathbf{u})=w_{k}^{i}(\mathbf{u})\text{, \ \ \ \
\ \ \ \ }i,k=1,2,...,N.  \label{ghm}
\end{equation}%
yields (in an implicit form) a general solution for (\textbf{\ref{2}}). This
is the \textit{generalized hodograph} method.

\textbf{Definition }(\textbf{\cite{Maks+Benney}}, \textbf{\cite{Maks+kdv}}, 
\textbf{\cite{Maks+Tsar}}): \textit{The semi-Hamiltonian hydrodynamic type
system }(\textbf{\ref{2}}) \textit{is said to be the Egorov, if} (\textbf{%
\ref{2}}) \textit{has the couple of conservation laws}%
\begin{equation}
\partial _{t}a=\partial _{x}b\text{, \ \ \ \ \ }\partial _{t}b=\partial
_{x}c.  \label{1}
\end{equation}

If (\textbf{\ref{2}}) is the Egorov hydrodynamic type system, then each
commuting flow (\textbf{\ref{com}}) has the similar pair of conservation
laws (\textbf{\ref{1}}) (see details in \textbf{\cite{Maks+Benney}}, \textbf{%
\cite{Maks+kdv}}, \textbf{\cite{Maks+Tsar}})%
\begin{equation}
\partial _{\tau }a=\partial _{x}h\text{, \ \ \ \ \ \ \ }\partial _{\tau
}h=\partial _{x}f.  \label{con}
\end{equation}

In this paper we describe a very important class of conservation laws -- the
Egorov hydrodynamic type systems (see \textbf{\cite{Dubr}}, \textbf{\cite%
{Gibbons}}, \textbf{\cite{Gib+Kod}}, \textbf{\cite{Krich}}, \textbf{\cite%
{Manas}}, \textbf{\cite{Maks+Benney}}, \textbf{\cite{Maks+kdv}}, \textbf{%
\cite{Maks+Tsar}}, \textbf{\cite{Tsar}}, \textbf{\cite{Zakh}}), which have
plenty applications in different areas of pure and applied mathematics,
physics, biology and chemistry.

In this paper we introduce the\textit{\ \textbf{extended} hodograph method}
for the Egorov hydrodynamic type systems; we significantly improve the
generalized hodograph method (\textbf{\ref{ghm}}) for the Egorov
hydrodynamic type systems; if the Egorov hydrodynamic type systems possess
Hamiltonian structure, then $N$ series of conservation laws and commuting
flows can be found iteratively.

The paper is organized in the following order. In the second section the
Egorov hydrodynamic type systems are considered with the aid of the
differential geometry (conjugate curvilinear coordinate nets). The \textit{%
Egorov basic set} of conservation laws and commuting flows is introduced. In
the third section the generalized hodograph method is adopted for the Egorov
hydrodynamic type systems. In the fourth section the extended hodograph
method is presented. In the fifth section the integrability criterion of the
Egorov hydrodynamic type systems is given. In the sixth section orthogonal
curvilinear coordinate nets with symmetric rotation coefficients are
considered. In the seventh section the Egorov hydrodynamic type systems
possessing a local Hamiltonian structure are investigated. Corresponding
associativity equations are derived. In the eighth section reciprocal
transformations preserving local Hamiltonian structure are found. In the
ninth section the Egorov three orthogonal curvilinear coordinate nets are
considered. Transformations connecting the Egorov hydrodynamic type systems
associated with local Hamiltonian structures are described. In the tenth
section the Egorov hydrodynamic type systems associated with nonlocal
Hamiltonian structure are briefly mentioned.

\section{\textit{Conjugate} curvilinear coordinate nets}

The theory of \textit{conjugate curvilinear coordinate nets} is developed by
G. Darboux (see \textbf{\cite{Darboux}}). The linear system%
\begin{equation}
\partial _{i}H_{k}=\beta _{ik}H_{i}\text{, \ \ \ \ }i\neq k  \label{8}
\end{equation}%
is integrable, if (i.e. if $\partial _{j}(\partial _{i}H_{k})=\partial
_{i}(\partial _{j}H_{k})$, \ \ $i\neq j\neq k$)%
\begin{equation}
\partial _{i}\beta _{jk}=\beta _{ji}\beta _{ik}\text{, \ \ \ \ }i\neq j\neq
k.  \label{9}
\end{equation}%
The functions $\beta _{ik}$ are said to be \textit{the rotation coefficients
of conjugate curvilinear coordinate nets}, the functions $H_{i}$ are said to
be \textit{the Lame coefficients}.

Let us take some solution of the nonlinear PDE system (\textbf{\ref{9}}).
Suppose the general solution $H_{i}$ of the linear PDE system (\textbf{\ref%
{8}}) (parameterized by $N$ arbitrary functions of a single variable) is
found. Let us take any two arbitrary solutions $H_{(2)i}$ and $H_{(1)i}$ of (%
\textbf{\ref{8}}), then one can construct the integrable hydrodynamic type
system (see \textbf{\cite{Tsar}})%
\begin{equation}
r_{t^{2}}^{i}=\frac{H_{(2)i}}{H_{(1)i}}r_{t^{1}}^{i},  \label{10}
\end{equation}%
where the Riemann invariants $r^{i}$ are implicit functions of ``times'' $%
t^{1}$ and $t^{2}$, which usually are called as $x$ and $t$, respectively.
The general solution of this hydrodynamic type system is given by the
generalized hodograph method (see \textbf{\cite{Tsar}}):

\textbf{Theorem} \textbf{\cite{Tsar}}: \textit{The algebraic system}%
\begin{equation}
xH_{(1)i}+tH_{(2)i}=H_{i}\text{, \ \ \ \ }i=1,2,...,N  \label{11}
\end{equation}%
\textit{yields a general solution (in an implicit form) of the hydrodynamic
type system }(\textbf{\ref{10}}).

Thus, the general solution of the nonlinear PDE system (\textbf{\ref{9}})
together with the general solution of the linear PDE system (\textbf{\ref{8}}%
) describe all possible semi-Hamiltonian hydrodynamic type systems (\textbf{%
\ref{10}}).

Taking arbitrary solutions $\psi _{i}^{(\beta )}$ of the \textit{adjoint}
linear system (see (\textbf{\ref{8}}))%
\begin{equation}
\partial _{i}\psi _{k}=\beta _{ki}\psi _{i}\text{, \ \ \ \ }i=1\text{, }2%
\text{, . . . , }N,  \label{12}
\end{equation}%
one can construct conservation laws written in the potential form%
\begin{equation}
dz^{\beta }=\underset{\gamma }{\sum }a_{\gamma }^{\beta }(\mathbf{r}%
)dt^{\gamma }  \label{14}
\end{equation}%
for the hydrodynamic type systems%
\begin{equation}
r_{t^{\beta }}^{i}=\frac{H_{(\beta )i}}{H_{(\gamma )i}}r_{t^{\gamma }}^{i},%
\text{ \ \ \ }i=1\text{, }2\text{, . . . , }N,  \label{13}
\end{equation}%
where%
\begin{equation}
\partial _{i}a_{\gamma }^{\beta }=\psi _{i}^{(\beta )}H_{(\gamma )i}.
\label{15}
\end{equation}%
It means, for example, that the hydrodynamic type system (\textbf{\ref{10}})
has an infinite number of the conservation laws%
\begin{equation*}
\partial _{t^{2}}a_{1}^{\beta }=\partial _{t^{1}}a_{2}^{\beta },
\end{equation*}%
parameterized by $N$ arbitrary functions of a single variable.

Let us consider the hydrodynamic type system (\textbf{\ref{10}}) together
with its $M-2$ nontrivial commuting flow (see (\textbf{\ref{13}}))%
\begin{equation}
r_{t^{m}}^{i}=\frac{H_{(m)i}}{H_{(1)i}}r_{t^{1}}^{i}\text{, \ \ \ \ \ \ \ }%
m=2,3,...,N.  \label{sym}
\end{equation}%
Then the generalized hodograph method (see (\textbf{\ref{11}})) yields the
general solution (see \textbf{\cite{Gib+Kod}})%
\begin{equation}
xH_{(1)i}+tH_{(2)i}+\overset{N}{\underset{k=3}{\sum }}t^{k}H_{(k)i}=H_{i},
\label{22}
\end{equation}%
where $N$ Riemann invariants $r^{i}$ are functions of $N$ independent
variables $t^{k}$. This is invertible transformation $%
r^{i}(t^{1},t^{2},...,t^{N})$.

In this article we restrict a consideration of the hydrodynamic type systems
(\textbf{\ref{10}}) (see also (\textbf{\ref{13}}) and (\textbf{\ref{sym}}))
on the symmetric case%
\begin{equation}
\beta _{ik}=\beta _{ki}\text{, \ \ \ \ }i\neq k.  \label{16}
\end{equation}%
Corresponding conjugate curvilinear coordinate nets (\textbf{\ref{9}}) were
introduced by G. Darboux in 1866 and investigated by D.Th. Egorov in 1901 in
his thesis (see \textbf{\cite{Egorov}}). It was G. Darboux (see \textbf{\cite%
{Darboux}}) who proposed to call them (see (\textbf{\ref{9}}) and (\textbf{%
\ref{16}})) the Egorov curvilinear coordinate systems. From the point of
view of integrability properties a remarkable progress was achieved by L.
Bianchi in 1915 (see \textbf{\cite{Bianchi}}). He found a B\"{a}cklund
transformation for this problem and established the permutability property
as well as the superposition formula for it in the flat case, i.e. when
conjugate curvilinear coordinate net becomes to be orthogonal%
\begin{equation}
\partial _{i}\beta _{ik}+\partial _{k}\beta _{ki}+\underset{m\neq i,k}{\sum }%
\beta _{mi}\beta _{mk}=0.  \label{17}
\end{equation}%
Namely, the Egorov orthogonal curvilinear coordinate nets were extensively
investigated at that time.

If rotation coefficients $\beta _{ik}$ are symmetric, then the linear
problems (\textbf{\ref{8}}) and (\textbf{\ref{12}}) coincide, and
corresponding conservation laws (see (\textbf{\ref{14}}) and (\textbf{\ref%
{15}})) can be taken in the symmetric form too%
\begin{equation}
d\Omega _{\beta }=\underset{\gamma }{\sum }a_{\beta \gamma }dt^{\gamma },
\label{18}
\end{equation}%
where%
\begin{equation}
\partial _{i}a_{\beta \gamma }=H_{(\beta )i}H_{(\gamma )i}\text{.}
\label{19}
\end{equation}%
It means that one can introduce the function $\Omega $ determined by its
second derivatives%
\begin{equation*}
a_{\beta \gamma }=\frac{\partial ^{2}\Omega }{\partial t^{\beta }\partial
t^{\gamma }}\text{, \ \ \ \ \ \ \ }\Omega _{\beta }=\frac{\partial \Omega }{%
\partial t^{\beta }}.
\end{equation*}%
Then the commuting flows (\textbf{\ref{sym}}) are determined by the unique
function $\Omega $. Thus, the hydrodynamic type system (\textbf{\ref{10}})
has the Egorov couple of conservation laws (\textbf{\ref{1}}) (see (\textbf{%
\ref{19}})), where%
\begin{equation}
\partial _{i}a_{11}=H_{(1)i}^{2}\text{, \ \ \ \ }\partial
_{i}a_{12}=H_{(2)i}H_{(1)i}\text{, \ \ \ \ }\partial _{i}a_{22}=H_{(2)i}^{2},
\label{21}
\end{equation}

\textbf{Definition }(\textbf{\cite{Maks+Benney}}, \textbf{\cite{Maks+kdv}}, 
\textbf{\cite{Maks+Tsar}}): \textit{The conservation law density }$a_{11}$ 
\textit{is said to be the \textbf{potential} of the Egorov metric}.

Let us take $N$ particular solutions $H_{(\beta )i}$ of the linear system (%
\textbf{\ref{8}}). If the rotation coefficients $\beta _{ik}$ are symmetric
(see (\textbf{\ref{16}})), then the Egorov hydrodynamic type systems (%
\textbf{\ref{sym}}) can be written together in the symmetric potential form (%
\textbf{\ref{18}}).

In the next section we improve the generalized hodograph method (see (%
\textbf{\ref{11}}), (\textbf{\ref{22}})) for the Egorov hydrodynamic type
systems.

\section{The \textit{generalized} hodograph method}

Let us introduce $N$ field variables $a_{k}$ such that $\partial
_{i}a_{k}=H_{(k)i}H_{(1)i}$ (i.e. $a_{k}\equiv a_{k1}$, see (\textbf{\ref{18}%
}) and (\textbf{\ref{19}})).

Then the algebraic system (\textbf{\ref{22}}) for the Egorov hydrodynamic
type systems (\textbf{\ref{sym}})%
\begin{equation}
\partial _{t^{k}}a_{n}=\partial _{t^{1}}a_{kn}(\mathbf{a})  \label{sima}
\end{equation}%
can be written in the form%
\begin{equation}
\overset{N}{\underset{k=1}{\sum }}t^{k}\partial _{i}a_{k}=\partial _{i}h%
\text{ \ \ \ \ \ }\Leftrightarrow \text{\ \ \ \ \ }\overset{N}{\underset{k=1}%
{\sum }}t^{k}da_{k}=dh,  \label{godo}
\end{equation}%
where the conservation law density $h(a_{1},a_{2},...,a_{N})$ is determined
by its derivatives $\partial _{i}h=H_{i}H_{(1)i}$ (see (\textbf{\ref{19}}))
and parameterized by $N$ arbitrary functions of a single variable. The
fluxes $p_{k}(\mathbf{a})$ of this conservation law (cf. (\textbf{\ref{cons}}%
))%
\begin{equation}
\partial _{t^{k}}h(\mathbf{a})=\partial _{t^{1}}p_{k}(\mathbf{a})
\label{ful}
\end{equation}%
can be found in quadratures%
\begin{equation*}
dp_{k}(\mathbf{a})=\frac{\partial h(\mathbf{a})}{\partial a_{n}}da_{kn}(%
\mathbf{a}).
\end{equation*}

\textbf{Lemma}: \textit{The Egorov hydrodynamic type systems} (\textbf{\ref%
{sima}}) \textit{possess an arbitrary commuting flow (see} (\textbf{\ref{com}%
}) \textit{and} (\textbf{\ref{con}}))%
\begin{equation}
\partial _{\tau }a_{1}=\partial _{t^{1}}h(\mathbf{a})\text{, \ \ \ \ \ \ \ \ 
}\partial _{\tau }a_{k}=\partial _{t^{1}}p_{k}(\mathbf{a})\text{, \ \ \ \ }%
k=2,3,...,N.  \label{lem}
\end{equation}

\textbf{Proof}: can be obtained from the compatibility condition $\partial
_{t^{k}}(\partial _{\tau }a_{1})=\partial _{\tau }(\partial _{t^{k}}a_{1})$.

Thus, we have the generalized hodograph method \textit{adopted} for the
Egorov hydrodynamic type systems.

\textbf{Theorem}: \textit{A general solution of the Egorov hydrodynamic type
systems} (\textbf{\ref{sym}}) \textit{is given by the algebraic system (in
an implicit form, see }(\textbf{\ref{godo}})\textit{)}%
\begin{equation}
t^{k}=\frac{\partial h(\mathbf{a})}{\partial a_{k}}\text{.}  \label{main}
\end{equation}

\textbf{Corollary}: The function $\Omega $ can be expressed explicitly via
the field variables $a_{k}$ and found in quadratures in two steps (see (%
\textbf{\ref{18}}) and below)%
\begin{equation*}
d\Omega _{k}=a_{ks}(\mathbf{a})d\frac{\partial h(\mathbf{a})}{\partial a_{k}}%
\text{, \ \ \ \ \ \ }d\Omega =\Omega _{k}(\mathbf{a})d\frac{\partial h(%
\mathbf{a})}{\partial a_{k}}.
\end{equation*}

In the next section we introduce the extended hodograph method for
integrability of the Egorov hydrodynamic type systems based on this
symmetric representation.

\section{The \textit{extended} hodograph method}

Any two-component system (\textbf{\ref{2}}) is integrable by the \textit{%
hodograph} method (see, for instance, \textbf{\cite{Yanenko}}). Let us adopt
this hodograph method for the Egorov hydrodynamic type systems. The
hydrodynamic type system (\textbf{\ref{2}}) can be written in the
conservative form (\textbf{\ref{1}}). This couple of conservation laws can
be written in the potential form%
\begin{equation*}
dy=adx+bdt\text{, \ \ \ \ \ \ \ }dz=bdx+c(a,b)dt.
\end{equation*}%
Using the Legendre transform $\Phi =ax+bt-y$, $\xi =bx+ct-z$ the above
couple of equations can be written in the form%
\begin{equation*}
d\Phi =xda+tdb\text{, \ \ \ \ \ \ \ \ }d\xi =xdb+tdc(a,b).
\end{equation*}%
Since $x_{b}=t_{a}$ (see the first above equation), then (see the second
above equation) the compatibility condition $(\xi _{a})_{b}=(\xi _{b})_{a}$%
\begin{equation}
(x+tc_{b})_{a}=(tc_{a})_{b}  \label{lin}
\end{equation}%
can be written as the \textit{linear} PDE equation of the second order%
\begin{equation}
\Phi _{aa}+c_{b}\Phi _{ab}=c_{a}\Phi _{bb}.  \label{fik}
\end{equation}%
Thus, the hodograph method is the transformation from the field variables
(unknown functions) $a(x,t)$ and $b(x,t)$ of the \textbf{quasilinear} system
(\textbf{\ref{2}}) to the new field variables (independent variables) $%
x(a,b) $ and $t(a,b)$ of the \textbf{linear} system (\textbf{\ref{lin}}).

\textbf{Remark}: Since (see (\textbf{\ref{main}})) $x=h_{a}$ and $t=h_{b}$,
then $\Phi \equiv h$ and $\xi \equiv p$. Thus, the above equation (\textbf{%
\ref{fik}}) coincides with the equation describing conservation law
densities $h(a,b)$ of the hydrodynamic type system $a_{t}=b_{x}$, $%
b_{t}=\partial _{x}c(a,b)$.

\textbf{Example}: the nonlinear elasticity equation (see, for instance, 
\textbf{\cite{Nutku}}) is determined by the function $c(a)$. The above
linear equation is reducible to the hypergeometric equation if $c(a)=a^{n}$, 
$c(a)=\ln a$, $c(a)=e^{a}$.

Let us consider the couple of three component Egorov hydrodynamic type system%
\begin{equation}
\begin{array}{ccccccccc}
a_{t}=b_{x}\text{,} &  &  &  &  &  &  &  & a_{y}=c_{x}\text{,} \\ 
&  &  &  &  &  &  &  &  \\ 
b_{t}=\partial _{x}u(a,b,c)\text{,} &  &  &  &  &  &  &  & b_{y}=\partial
_{x}v(a,b,c)\text{,} \\ 
&  &  &  &  &  &  &  &  \\ 
c_{t}=\partial _{x}v(a,b,c), &  &  &  &  &  &  &  & c_{y}=\partial
_{x}w(a,b,c),%
\end{array}
\label{eg}
\end{equation}%
which can be written in the potential symmetric form (\textbf{\ref{sima}})%
\begin{equation}
d\left( 
\begin{array}{c}
\xi ^{1} \\ 
\xi ^{2} \\ 
\xi ^{3}%
\end{array}%
\right) =\left( 
\begin{array}{ccc}
a & b & c \\ 
b & u & v \\ 
c & v & w%
\end{array}%
\right) d\left( 
\begin{array}{c}
t^{1} \\ 
t^{2} \\ 
t^{3}%
\end{array}%
\right) ,  \label{tri}
\end{equation}%
where $y=t^{3}$.

If functions $a(x,t,y)$, $b(x,t,y)$ and $c(x,t,y)$ are common for both
hydrodynamic type systems (\textbf{\ref{eg}}), then the compatibility
condition $\partial _{t}(\partial _{y}a)=\partial _{y}(\partial _{t}a)$ 
\textit{satisfies identically} and the compatibility condition $\partial
_{t}(\partial _{y}b)=\partial _{y}(\partial _{t}b)$ is \textit{identically
coincides} with the compatibility condition $\partial _{t}(\partial
_{y}c)=\partial _{y}(\partial _{t}c)$.

\textbf{Lemma}: \textit{If two hydrodynamic type systems} (\textbf{\ref{eg}}%
) \textit{commute, then the relationship between three coefficients} $%
u(a,b,c)$, $v(a,b,c)$ \textit{and} $w(a,b,c)$ \textit{is given by}%
\begin{eqnarray}
\partial _{b}\frac{u_{a}v_{b}+v_{a}v_{c}-u_{b}v_{a}}{u_{c}} &=&\partial _{a}%
\frac{v_{a}+v_{b}v_{c}}{u_{c}},  \notag \\
&&  \notag \\
\partial _{b}\frac{u_{c}v_{b}+v_{c}^{2}-u_{b}v_{c}-u_{a}}{u_{c}} &=&\partial
_{c}\frac{v_{a}+v_{b}v_{c}}{u_{c}},  \label{cc} \\
&&  \notag \\
\partial _{a}\frac{u_{c}v_{b}+v_{c}^{2}-u_{b}v_{c}-u_{a}}{u_{c}} &=&\partial
_{c}\frac{u_{a}v_{b}+v_{a}v_{c}-u_{b}v_{a}}{u_{c}},  \notag
\end{eqnarray}%
where%
\begin{equation}
dw=\frac{%
(u_{a}v_{b}+v_{a}v_{c}-u_{b}v_{a})da+(v_{a}+v_{b}v_{c})db+(u_{c}v_{b}+v_{c}^{2}-u_{b}v_{c}-u_{a})dc%
}{u_{c}}.  \label{j}
\end{equation}

Under the Legendre transformation $h=at^{1}+bt^{2}+ct^{3}-\xi ^{1}$, $%
p=bt^{1}+ut^{2}+vt^{3}-\xi ^{2}$, $q=ct^{1}+vt^{2}+wt^{3}-\xi ^{3}$, the
above symmetric system (\textbf{\ref{tri}}) reduces to the three
differentials%
\begin{equation*}
dh=t^{1}da+t^{2}db+t^{3}dc\text{, \ \ \ \ \ \ }dp=t^{1}db+t^{2}du+t^{3}dv%
\text{, \ \ \ \ \ \ \ }dq=t^{1}dc+t^{2}dv+t^{3}dw\text{.}
\end{equation*}%
Since (see the first above equation and (\textbf{\ref{main}})) $t^{1}=h_{a}$%
, $t^{2}=h_{b}$, $t^{3}=h_{c}$, then (see the second and third above
equations) the compatibility conditions ($(p_{a})_{b}=(p_{b})_{a}$, $%
(q_{a})_{b}=(q_{b})_{a}$, etc)%
\begin{eqnarray*}
(t^{2}u_{a}+t^{3}v_{a})_{b} &=&(t^{1}+t^{2}u_{b}+t^{3}v_{b})_{a}, \\
&& \\
(t^{2}u_{c}+t^{3}v_{c})_{a} &=&(t^{2}u_{a}+t^{3}v_{a})_{c}, \\
&& \\
(t^{2}u_{c}+t^{3}v_{c})_{b} &=&(t^{1}+t^{2}u_{b}+t^{3}v_{b})_{c}
\end{eqnarray*}%
can be written as the \textit{linear} PDE system (where we use the temporary
notation $\Phi =\xi ^{1}$)%
\begin{eqnarray}
u_{a}\Phi _{bb}+v_{a}\Phi _{bc} &=&\Phi _{aa}+u_{b}\Phi _{ab}+v_{b}\Phi
_{ac},  \notag \\
&&  \notag \\
u_{a}\Phi _{bc}+v_{a}\Phi _{cc} &=&u_{c}\Phi _{ab}+v_{c}\Phi _{ac},
\label{bb} \\
&&  \notag \\
u_{c}\Phi _{bb}+v_{c}\Phi _{bc} &=&\Phi _{ac}+u_{b}\Phi _{bc}+v_{b}\Phi
_{cc}.  \notag
\end{eqnarray}

Thus, the above described transformation from the field variables (unknown
functions) $a(x,t,y)$, $b(x,t,y)$ and $c(x,t,y)$ of the \textbf{quasilinear}
system (\textbf{\ref{eg}}) to the new field variables (independent
variables) $x(a,b,c)$, $t(a,b,c)$ and $y(a,b,c)$ of the \textbf{linear}
system (\textbf{\ref{bb}}) is nothing but the \textit{extended hodograph
method} for a couple of three component hydrodynamic type systems.

The compatibility conditions $(\Phi _{cc})_{b}=(\Phi _{bc})_{c}$, $(\Phi
_{cc})_{a}=(\Phi _{ac})_{c}$, $(\Phi _{ac})_{b}=(\Phi _{bc})_{a}$ of the
over-determined system (\textbf{\ref{bb}}) are equivalent (\textbf{\ref{cc}}%
).

\textbf{Remark}: Since (see (\textbf{\ref{main}})) $x=h_{a}$, $t=h_{b}$ and $%
y=h_{c}$, then (see (\textbf{\ref{ful}})) $\xi ^{1}\equiv h$, $\xi
^{1}\equiv p$ and $\xi ^{1}\equiv q$ (where we use temporary notation $%
F_{1}=p$ and $F_{2}=q$). Thus, the above system (\textbf{\ref{bb}})
coincides with the system describing conservation law densities $h(a,b,c)$
of both hydrodynamic type systems (\textbf{\ref{eg}}).

Obviously, this construction easily can be extended on $N$ component case.

\section{\textit{Natural} extra commuting flows. The integrability criterion}

Without lost of generality let us restrict our consideration on the first
nontrivial three component Egorov hydrodynamic type system written in the
conservative form%
\begin{equation}
a_{t}=b_{x}\text{, \ \ \ \ \ \ \ }b_{t}=\partial _{x}u(a,b,c)\text{, \ \ \ \
\ \ \ }c_{t}=\partial _{x}v(a,b,c).  \label{23}
\end{equation}%
If this hydrodynamic type system is semi-Hamiltonian, then it must admit the 
\textit{natural} commuting flow (cf. (\textbf{\ref{eg}}))%
\begin{equation}
a_{y}=c_{x}\text{, \ \ \ \ \ \ \ }b_{y}=\partial _{x}v(a,b,c)\text{, \ \ \ \
\ \ \ \ }c_{y}=\partial _{x}w(a,b,c).  \label{25}
\end{equation}

Indeed, since (\textbf{\ref{23}}) is semi-Hamiltonian, then this
hydrodynamic type system has an infinite series of conservation laws%
\begin{equation*}
\partial _{t}h(a,b,c)=\partial _{x}p(a,b,c)
\end{equation*}%
and commuting flows. Each of them can be written in the form (\textbf{\ref%
{lem}}) (see \textbf{\cite{Maks+Tsar}})%
\begin{equation}
a_{\tau }=\partial _{x}h(a,b,c)\text{, \ \ \ \ \ \ \ }b_{\tau }=\partial
_{x}p(a,b,c)\text{, \ \ \ \ \ \ \ \ }c_{\tau }=\partial _{x}q(a,b,c),
\label{u}
\end{equation}%
where the commuting flow (\textbf{\ref{25}}) has the conservation law%
\begin{equation*}
\partial _{y}h(a,b,c)=\partial _{x}q(a,b,c).
\end{equation*}%
The function $h(a,b,c)$ satisfies the linear PDE system of the first order
with variable coefficients. Thus, it is very difficult to find the function $%
h(a,b,c)$ in general case. However, at least three particular solutions $%
h(a,b,c)$ are given a priori (see (\textbf{\ref{23}})); i.e. $h_{(1)}=a$, $%
h_{(2)}=b$ and $h_{(3)}=c$. The first choice is trivial, the second choice
is given by (\textbf{\ref{23}}). Thus, we can try to reconstruct an extra
commuting flow, where the first conservation law is%
\begin{equation*}
a_{y}=c_{x}.
\end{equation*}%
Moreover, the compatibility condition $\partial _{y}(\partial
_{t}a)=\partial _{t}(\partial _{y}a)$ leads to the next conservation law%
\begin{equation*}
b_{y}=\partial _{x}v(a,b,c).
\end{equation*}%
Finally, the compatibility condition $\partial _{y}(\partial _{t}b)=\partial
_{t}(\partial _{y}b)$ leads to the criterion of an integrability for the
three component Egorov hydrodynamic type systems (\textbf{\ref{23}}).

\textbf{Criterion of integrability}: \textit{The Egorov hydrodynamic type
system} (\textbf{\ref{23}}) \textit{has the commuting flow} \textit{iff the
function} $w$ \textit{can be found in quadratures }(\textbf{\ref{j}}).

The compatibility condition $\partial _{y}(\partial _{t}c)=\partial
_{t}(\partial _{y}c)$ leads to the same result. Thus, if the compatibility
conditions are fulfilled, then the Egorov hydrodynamic type system (\textbf{%
\ref{23}}) has the extra commuting flow (\textbf{\ref{25}}).

\textbf{Remark}: Taking into account that $b=z_{t}=\xi _{x}$, $%
c=z_{y}=\sigma _{x}$ (see (\textbf{\ref{sima}})) a new potential function $%
\Omega $ can be introduced, where%
\begin{equation*}
d\Omega =zdx+\xi dt+\sigma dy\text{.}
\end{equation*}%
Then, the couple of hydrodynamic type system (\textbf{\ref{23}}), (\textbf{%
\ref{25}}) can be written in the form%
\begin{equation*}
\Omega _{tt}=u(\Omega _{xx},\Omega _{xt},\Omega _{xy})\text{, \ \ \ \ \ \ }%
\Omega _{yt}=v(\Omega _{xx},\Omega _{xt},\Omega _{xy})\text{, \ \ \ \ \ \ }%
\Omega _{yy}=w(\Omega _{xx},\Omega _{xt},\Omega _{xy}).
\end{equation*}%
These equations separately can be considered as 2+1 quasilinear equations.
However, in general case all of them are non-integrable (see, for instance, 
\textbf{\cite{FerKarMax}}). In some cases one of them can be integrable (by
the method of hydrodynamic reductions; \textbf{\cite{FerKarMax}}). Then two
other equations are reductions of higher order commuting flows on a three
component case. These three 2+1 equations are compatible if (\textbf{\ref{cc}%
}) are fulfilled.

Obviously, the same integrability criterion can be derived for $N$ component
hydrodynamic type system written in the conservative form and containing (%
\textbf{\ref{1}}).

\textbf{Theorem}: \textit{If the Egorov hydrodynamic type system (see} (%
\textbf{\ref{2}}) \textit{and} (\textbf{\ref{1}})) \textit{is
semi-Hamiltonian, then} $M-2$ \textit{nontrivial commuting flows can be
found in quadratures. In such a case these hydrodynamic type systems can be
written in the symmetric form} (\textbf{\ref{sima}}).

\textbf{Proof}: Consider the Egorov hydrodynamic type system written in the
form%
\begin{equation}
\partial _{t}a_{1}=\partial _{x}a_{2}\text{, \ \ \ \ \ }\partial
_{t}a_{k}=\partial _{x}b_{k}(\mathbf{a})\text{, \ \ \ \ }k=2,3,...,N.
\label{egor}
\end{equation}%
If this hydrodynamic type system is semi-Hamiltonian, then $(N-1)\times
(N-1) $ symmetric matrix with the elements $a_{kn}(\mathbf{a})$ can be
introduced, where $a_{k2}(\mathbf{a})=b_{k}(\mathbf{a})$ and all other
elements are not determined yet. Since (\textbf{\ref{egor}}) is
semi-Hamiltonian, then $a_{1}$ is the potential of the Egorov metric (see (%
\textbf{\ref{10}}) and (\textbf{\ref{21}})), i.e. $\partial
_{i}a_{1}=H_{(1)i}^{2}$. Then (see (\textbf{\ref{10}})) $\partial
_{i}a_{k}=H_{(k)i}H_{(1)i}$ and $\partial _{i}b_{k}=H_{(k)i}H_{(2)i}$. Then,
indeed, all other components $a_{kn}$ can be found in quadratures (see (%
\textbf{\ref{19}}))%
\begin{equation*}
da_{kn}=\sum H_{(k)m}H_{(n)m}dr^{m}=\sum \frac{\partial _{m}a_{k}\partial
_{m}a_{n}}{\partial _{m}a_{1}}dr^{m}.
\end{equation*}

\textbf{Definition}: \textit{The commuting flows} (\textbf{\ref{sym}}) 
\textit{written in the conservative form }(\textbf{\ref{sima}}) \textit{are
said to be the \textbf{Egorov basic set}}.

\textbf{Remark}: The above proof can be obtained without the Riemann
invariants $r^{k}$. The compatibility conditions $\partial _{t^{k}}(\partial
_{t^{n}}a_{m})=\partial _{t^{n}}(\partial _{t^{k}}a_{m})$ lead to the full
set of relationships between the coefficients $a_{kn}$, which are
complicated in the variables $a^{k}$ (see, for instance, the above case (%
\textbf{\ref{j}})).

The main reason for consideration of local Hamiltonian structures for the
Egorov hydrodynamic type systems (\textbf{\ref{23}}) is following: the
general solution (\textbf{\ref{main}}) is determined by the general solution
(\textbf{\ref{bb}}) (in the three component case), which is parameterized by
three arbitrary functions of a single variable. However, in general case
this linear PDE system has variable coefficients. The Hamiltonian structure
leads to a reduction of (\textbf{\ref{bb}}) to the linear ODE system, whose
coefficients $h^{(k)}$ can be found recursively (see below).

\section{\textit{Orthogonal} curvilinear coordinate nets}

Local Hamiltonian structures of hydrodynamic type systems integrable by the
generalized hodograph method are connected with the theory of \textit{%
orthogonal curvilinear coordinate nets} (see \textbf{\cite{Darboux}}). The
zero curvature condition (\textbf{\ref{17}}) is a consequence of
relationship of two linear problems (\textbf{\ref{8}}) and (\textbf{\ref{12}}%
)%
\begin{equation}
H_{i}=\partial _{i}\psi _{i}+\underset{m\neq i}{\sum }\beta _{mi}\psi _{m}.
\label{connect}
\end{equation}%
The existence of this first order transformation is equivalent (see \textbf{%
\cite{Tsar}}) the existence of local Hamiltonian structure for corresponding
hydrodynamic type system (\textbf{\ref{10}})%
\begin{equation}
a_{t}^{i}=\partial _{x}\left( \bar{g}^{ik}\frac{\delta \mathbf{h}}{\delta
a^{k}}\right) ,  \label{ham}
\end{equation}%
where the Hamiltonian is $\mathbf{h}=\int h(\mathbf{a})dx$, $\partial
_{i}h=\psi _{i}^{(2)}H_{(1)i}$, the momentum density is $P=\bar{g}%
_{ik}a^{i}a^{k}/2$, $\partial _{i}P=\psi _{i}^{(1)}H_{(1)i}$, where 
\begin{equation*}
H_{(2)i}=\partial _{i}\psi _{i}^{(2)}+\underset{m\neq i}{\sum }\beta
_{mi}\psi _{m}^{(2)}\text{, \ \ \ \ \ \ \ \ \ }H_{(1)i}=\partial _{i}\psi
_{i}^{(1)}+\underset{m\neq i}{\sum }\beta _{mi}\psi _{m}^{(1)}
\end{equation*}%
and the flat coordinates $a^{k}$ are determined by its derivatives $\partial
_{i}a^{k}=\bar{\psi}_{i}^{(k)}H_{(1)i}$, the \textbf{constant} \textit{%
non-degenerate} metric%
\begin{equation}
\bar{g}^{ik}=\Sigma \bar{\psi}_{m}^{(i)}\bar{\psi}_{m}^{(k)}  \label{metr}
\end{equation}%
is given by%
\begin{equation}
0=\partial _{i}\bar{\psi}_{i}^{(k)}+\underset{m\neq i}{\sum }\beta _{mi}\bar{%
\psi}_{m}^{(k)}\text{, \ \ \ \ \ \ \ }k=1,2,...,N.  \label{nul}
\end{equation}

If the rotation coefficients $\beta _{ik}$ are symmetric (see (\textbf{\ref%
{16}})) then all above formulas simplify (see \textbf{\cite{Tsar}}). For
instance, the zero curvature condition (\textbf{\ref{17}}) reduces to $%
\delta \beta _{ik}=0$, where $\delta =\Sigma \partial _{m}$ is a shift
operator. It means, that the rotation coefficients $\beta _{ik}$ depend only
on the differences of the Riemann invariants $r^{n}-r^{m}$. The
corresponding linear transformation (\textbf{\ref{connect}}) reduces to $%
H_{i}=\delta \tilde{H}_{i}$ (where $H_{i}$ and $\tilde{H}_{i}$ are solutions
of the linear system (\textbf{\ref{8}})); the linear system (\textbf{\ref%
{nul}}) reduces to $\delta \bar{H}_{i}^{(k)}=0$. It means, that the \textit{%
basic} Lame coefficients $\bar{H}_{i}^{(k)}$ depend only on the differences
of the Riemann invariants $r^{n}-r^{m}$. Since the linear problems (\textbf{%
\ref{8}}) and (\textbf{\ref{12}}) coincide, then we are able to introduce
the Lame coefficients with up/sub-indexes%
\begin{equation*}
\bar{H}_{i}^{(k)}=\bar{g}^{ks}\bar{H}_{(s)i}\text{, \ \ \ \ \ \ \ \ \ }\bar{H%
}_{(k)i}=\bar{g}_{ks}\bar{H}_{i}^{(s)}\text{,}
\end{equation*}%
where the constant \textit{non-degenerate} metric is given by (cf. (\textbf{%
\ref{metr}}))%
\begin{equation*}
\bar{g}^{ik}=\sum \bar{H}^{(i)m}\bar{H}^{(k)m}\text{, \ \ \ \ \ \ \ \ }\bar{g%
}_{ik}=\sum \bar{H}_{(i)m}\bar{H}_{(k)m}.
\end{equation*}

Let us introduce the \textit{adjoint} flat coordinates $a_{i}=\bar{g}%
_{ik}a^{k}$. Then the Hamiltonian hydrodynamic type systems (\textbf{\ref%
{ham}}) can be written in the form%
\begin{equation}
\partial _{t}a_{i}=\partial _{x}\left( \bar{g}_{ik}\frac{\delta \mathbf{h}}{%
\delta a_{k}}\right) .  \label{zima}
\end{equation}

In this paper we restrict our consideration on two cases, where the the
Egorov hydrodynamic type system is written via flat coordinates $a^{k}$ (see
(\textbf{\ref{ham}})). The first case is determined by the condition that $a$
(see (\textbf{\ref{1}})) is a flat coordinate $a_{1}$ and $b$ is a flat
coordinate $a_{2}$. In general case $a$ is an arbitrary conservation law
density $h$.

\section{Local Hamiltonian structures. The complete integrability}

Without lost of generality for simplicity we restrict our consideration on
three component Egorov hydrodynamic type systems (\textbf{\ref{ham}}). Since
the symmetric constant matrix $\bar{g}^{ik}$ under a linear transformation
of independent variables can be reduced to the diagonal or skew-diagonal
case, we restrict our consideration of the first case ($a$ and $b$ are flat
coordinates) on these two sub-cases.

\textbf{1.} Let us consider the Egorov hydrodynamic type system (\textbf{\ref%
{23}}) with the local Hamiltonian structure 
\begin{equation}
a_{t}=\partial _{x}\frac{\partial h_{1}}{\partial c}\text{, \ \ \ }%
b_{t}=\partial _{x}\frac{\partial h_{1}}{\partial b}\text{, \ \ \ }%
c_{t}=\partial _{x}\frac{\partial h_{1}}{\partial a}\text{,}  \label{ega}
\end{equation}%
where $\partial h_{1}/\partial c=b$ (see (\textbf{\ref{1}})). Then we are
able to choose the extra commuting flow (\textbf{\ref{25}}) written in the
same Hamiltonian form%
\begin{equation}
a_{y}=\partial _{x}\frac{\partial h_{2}}{\partial c}\text{, \ \ \ }%
b_{y}=\partial _{x}\frac{\partial h_{2}}{\partial b}\text{, \ \ \ }%
c_{y}=\partial _{x}\frac{\partial h_{2}}{\partial a}\text{,}  \label{ego}
\end{equation}%
where $\partial h_{2}/\partial c=c$ (see (\textbf{\ref{eg}})).

\textbf{Definition}: \textit{The Egorov basic set of commuting flows} (%
\textbf{\ref{sima}}) \textit{is said to be \textbf{canonical} if the first
flat coordinate} $a_{1}$ \textit{is a potential of the Egorov metric, and
all other flat coordinates }$a_{k}$\textit{\ are fluxes of the first
conservation law} $\partial _{t^{k}}a_{1}=\partial _{t^{1}}a_{k}$, \textit{%
where the corresponding local Hamiltonian structure is given by} (\textbf{%
\ref{zima}}).

Since $\partial h_{2}/\partial b=\partial h_{1}/\partial a$, both Egorov
hydrodynamic type systems are determined by the sole function $z(a,b)$
satisfying the famous \textit{associativity equation }(see \textbf{\cite%
{Dubr}}, \textbf{\cite{Fer+centre}}, \textbf{\cite{Yavuz}})%
\begin{equation}
z_{aaa}=z_{abb}^{2}-z_{aab}z_{bbb},  \label{ace}
\end{equation}%
where $h_{1}=bc+z_{b}$, $h_{2}=c^{2}/2+z_{a}$ and the momentum density $P$
is given by $h_{0}=ac+b^{2}/2$. Since the flux $h$ of the potential $a$ (see
(\textbf{\ref{1}})) is a conservation law density for an arbitrary commuting
flow (see (\textbf{\ref{con}})), then the shift operator $\partial /\partial
c$ is consistent with linear system (\textbf{\ref{bb}}) describing
conservation law densities (i.e. $\partial \tilde{h}/\partial c=h$, where $h$
and $\tilde{h}$ are solutions of the linear system (\textbf{\ref{bb}})). It
is easy to understand, if to take into account that the first equation of an
arbitrary commuting flow is written in the form (cf. (\textbf{\ref{1}}) and (%
\textbf{\ref{ega}}))%
\begin{equation}
a_{\tau }=\partial _{x}\frac{\partial \tilde{h}}{\partial c}=\partial _{x}h.
\label{symmetry}
\end{equation}

\textbf{Theorem}: \textit{The Egorov hydrodynamic type system} (\textbf{\ref%
{ega}}) \textit{has three infinite series of conservation law densities,
whose coefficients} $h_{k}^{(n)}$ ($k=0,1,2$ and $n=0,1,2,...$) \textit{can
be found iteratively}.

\textbf{Proof}: Taking into account (\textbf{\ref{u}}) two other equations
of an arbitrary commuting flow are written in the form%
\begin{equation*}
b_{\tau }=\partial _{x}\frac{\partial \tilde{h}}{\partial b}=\partial _{x}p%
\text{, \ \ \ \ \ \ \ \ \ \ }c_{\tau }=\partial _{x}\frac{\partial \tilde{h}%
}{\partial a}=\partial _{x}q.
\end{equation*}%
Thus (see also (\textbf{\ref{ful}})),%
\begin{eqnarray}
dh_{n}^{(k+1)} &=&h_{n}^{(k)}dc+p_{n}^{(k)}db+q_{n}^{(k)}da,  \notag \\
&&  \notag \\
dp_{n}^{(k+1)} &=&p_{n}^{(k)}dc+\left(
q_{n}^{(k)}+z_{bbb}p_{n}^{(k)}+z_{abb}h_{n}^{(k)}\right) db+\left(
z_{abb}p_{n}^{(k)}+z_{aab}h_{n}^{(k)}\right) da,  \label{h} \\
&&  \notag \\
dq_{n}^{(k+1)} &=&q_{n}^{(k)}dc+\left(
z_{abb}p_{n}^{(k)}+z_{aab}h_{n}^{(k)}\right) db+\left(
z_{aab}p_{n}^{(k)}+z_{aaa}h_{n}^{(k)}\right) da,  \notag
\end{eqnarray}%
where $h_{0}^{(0)}=a$, $p_{0}^{(0)}=b$, $q_{0}^{(0)}=c$; $h_{1}^{(0)}=b$, $%
p_{1}^{(0)}=c+z_{bb}$, $q_{1}^{(0)}=z_{ab}$; $h_{2}^{(0)}=c$, $%
p_{2}^{(0)}=z_{ab}$, $q_{2}^{(0)}=z_{aa}$. Then three infinite series of
particular solutions by the generalized hodograph method (see (\textbf{\ref%
{main}})) are given in an implicit form (cf. (\textbf{\ref{u}}))%
\begin{equation}
x=q(a,b,c)\text{, \ \ \ }t=p(a,b,c)\text{, \ \ \ \ }y=h(a,b,c),  \label{trik}
\end{equation}%
where%
\begin{equation*}
h=\overset{2}{\underset{k=0}{\sum }}\overset{\infty }{\underset{n=0}{\sum }}%
\sigma _{kn}h_{k}^{(n)}\text{, \ \ \ \ \ \ \ \ }p=\overset{2}{\underset{k=0}{%
\sum }}\overset{\infty }{\underset{n=0}{\sum }}\sigma _{kn}p_{k}^{(n)}\text{%
, \ \ \ \ \ \ \ }q=\overset{2}{\underset{k=0}{\sum }}\overset{\infty }{%
\underset{n=0}{\sum }}\sigma _{kn}q_{k}^{(n)}
\end{equation*}%
and $\sigma _{kn}$ are arbitrary constants.

\textbf{Complete integrability}: In general $N$ component case the Egorov
basic set (\textbf{\ref{sima}}) is canonical (see (\textbf{\ref{ham}}) and (%
\textbf{\ref{zima}}))%
\begin{equation}
\partial _{t^{k}}a_{n}=\partial _{t^{1}}\left( \frac{\partial h_{k}}{%
\partial a^{n}}\right) =\partial _{t^{1}}a_{kn}(\mathbf{a}).  \label{beg}
\end{equation}%
It means, that%
\begin{equation*}
h_{k}=\frac{\partial F}{\partial a^{k}}.
\end{equation*}%
The compatibility conditions $\partial _{t^{k}}(\partial
_{t^{m}}a_{n})=\partial _{t^{m}}(\partial _{t^{k}}a_{n})$ lead to the WDVV
equation (see \textbf{\cite{Dubr}})%
\begin{equation*}
\frac{\partial ^{3}F}{\partial a^{k}\partial a^{i}\partial a^{s}}\bar{g}^{sp}%
\frac{\partial ^{3}F}{\partial a^{p}\partial a^{j}\partial a^{n}}=\frac{%
\partial ^{3}F}{\partial a^{j}\partial a^{i}\partial a^{s}}\bar{g}^{sp}\frac{%
\partial ^{3}F}{\partial a^{p}\partial a^{k}\partial a^{n}}.
\end{equation*}%
The canonical Egorov basic set must have $N$ infinite series of conservation
laws%
\begin{equation}
\partial _{t^{k}}h_{s}^{(p)}=\partial _{t^{1}}q_{k,s}^{(p)}\text{, \ \ \ }%
k,s=1,2,...,N\text{, \ \ \ }p=0,1,2,...  \label{bum}
\end{equation}%
and commuting flows (see \textbf{\cite{Tsar}})%
\begin{equation}
\partial _{t_{p}^{s}}a_{k}=\partial _{t^{1}}\left( \frac{\partial h_{s}^{(p)}%
}{\partial a^{k}}\right) \text{, \ \ \ }k,s=1,2,...,N\text{, \ \ \ }%
p=0,1,2,...  \label{big}
\end{equation}%
Since $a_{1}$ is a potential of the Egorov metric, then (see (\textbf{\ref%
{con}}))%
\begin{equation}
\partial _{t_{p}^{s}}a_{1}=\partial _{t^{1}}\left( \frac{\partial h_{s}^{(p)}%
}{\partial a^{1}}\right) =\partial _{t^{1}}h_{s}^{(p-1)},  \label{bag}
\end{equation}%
where $h_{s}^{(0)}\equiv a_{s}$ and $t_{(1)}^{s}\equiv t^{s}$, $s=1,2,...,N$%
. Thus, $\partial h_{s}^{(p)}/\partial a^{1}=h_{s}^{(p)}$. The compatibility
conditions $\partial _{t^{k}}(\partial _{t_{p}^{s}}a_{1})=\partial
_{t_{p}^{s}}(\partial _{t^{k}}a_{1})$ lead to (see (\textbf{\ref{bum}}) and (%
\textbf{\ref{big}}))%
\begin{equation}
\partial _{t^{k}}h_{s}^{(p-1)}=\partial _{t^{1}}\frac{\partial h_{s}^{(p)}}{%
\partial a^{k}}=\partial _{t^{1}}q_{k,s}^{(p-1)}.  \label{bug}
\end{equation}%
Thus, the conservation law densities $h_{s}^{(p+1)}$ can be found in
quadratures (see (\textbf{\ref{bag}}) and (\textbf{\ref{bug}}))%
\begin{equation}
dh_{s}^{(p+1)}=q_{k,s}^{(p)}da^{k}\text{, \ \ \ \ }s,k=1,2,...,N\text{, \ }%
p=0,1,2,...,  \label{sec}
\end{equation}%
where $q_{1,s}^{(p)}\equiv h_{s}^{(p)}$. The consistency of the conservation
laws (\textbf{\ref{bug}}) with (\textbf{\ref{beg}}) lead to the relationships%
\begin{equation}
\frac{\partial q_{k,s}^{(p)}}{\partial a^{i}}=\frac{\partial h_{s}^{(p)}}{%
\partial a^{n}}\bar{g}^{nm}\frac{\partial ^{3}F}{\partial a^{m}\partial
a^{k}\partial a^{i}}.  \label{dub}
\end{equation}%
The substitution of (\textbf{\ref{sec}}) in r.h.s. of (\textbf{\ref{dub}})
implies the recursion relationship%
\begin{equation*}
\frac{\partial ^{2}h_{k,s}^{(p+1)}}{\partial a^{i}\partial a^{k}}=\frac{%
\partial ^{3}F}{\partial a^{i}\partial a^{k}\partial a^{m}}\bar{g}^{mn}\frac{%
\partial h_{s}^{(p)}}{\partial a^{n}}\text{, \ \ \ \ \ }n=0,1,2,...,
\end{equation*}%
found by B.A. Dubrovin in \textbf{\cite{Dubr}}. Taking into account (\textbf{%
\ref{sec}}) the above formula leads to the iterative procedure (cf. (\textbf{%
\ref{h}}))%
\begin{equation*}
dq_{k,s}^{(p+1)}=q_{n,s}^{(p)}\bar{g}^{nm}\frac{\partial ^{3}F}{\partial
a^{m}\partial a^{k}\partial a^{i}}da^{i}\text{, \ \ \ \ }i,k,s,m,n=1,2,...,N%
\text{, \ }p=0,1,2,...,
\end{equation*}%
including (\textbf{\ref{sec}}). The complete integrability of the Egorov
Hamiltonian hydrodynamic type systems (\textbf{\ref{beg}}) was proved in 
\textbf{\cite{Tsar}}. Then the general solution can be given in implicit
form by the generalized hodograph method (see (\textbf{\ref{main}}) and cf. (%
\textbf{\ref{trik}}))%
\begin{equation*}
t^{k}=\bar{g}^{kn}\frac{\partial h}{\partial a^{n}}\text{, \ \ \ \ \ \ \ }h=%
\overset{N}{\underset{s=0}{\sum }}\overset{\infty }{\underset{p=0}{\sum }}%
\sigma _{ps}h_{s}^{(p)}\text{,}
\end{equation*}%
where $\sigma _{ps}$ are appropriate constants.

\textbf{Example}: The first choice $h_{0}^{(0)}=a$, $p_{0}^{(0)}=b$, $%
q_{0}^{(0)}=c$ determines the Egorov hydrodynamic type system%
\begin{equation*}
\partial _{\tau _{0}^{(0)}}a=\partial _{x}\left( ac+\frac{b^{2}}{2}\right) 
\text{, \ \ }\partial _{\tau _{0}^{(0)}}b=\partial _{x}\left(
bc+bz_{bb}+az_{ab}-z_{b}\right) \text{, \ \ }\partial _{\tau
_{0}^{(0)}}c=\partial _{x}\left( \frac{c^{2}}{2}+bz_{ab}+az_{aa}-z_{a}%
\right) .
\end{equation*}%
Thus, first four conservation laws for the canonical Egorov basic set (%
\textbf{\ref{ega}}), (\textbf{\ref{ego}}) together with the above commuting
flow can be written in the potential symmetric form%
\begin{equation}
d\left( 
\begin{array}{c}
\xi ^{1} \\ 
\xi ^{2} \\ 
\xi ^{3} \\ 
\xi ^{4}%
\end{array}%
\right) =\left( 
\begin{array}{cccc}
a & b & c & P \\ 
b & u & v & R \\ 
c & v & w & S \\ 
P & R & S & Q%
\end{array}%
\right) d\left( 
\begin{array}{c}
x \\ 
t \\ 
y \\ 
\tau%
\end{array}%
\right) ,  \label{tima}
\end{equation}%
where all coefficients can be found in quadratures (see (\textbf{\ref{h}}))%
\begin{eqnarray}
u &=&c+z_{bb}\text{, \ \ \ \ \ \ \ \ \ \ \ \ \ \ \ \ }v=z_{ab}\text{, \ \ \
\ \ \ \ \ \ \ \ \ \ \ \ \ \ }w=z_{aa},  \notag \\
&&  \notag \\
R &=&bc+bz_{bb}+az_{ab}-z_{b},\text{ \ \ \ \ \ \ \ }S=\frac{c^{2}}{2}%
+bz_{ab}+az_{aa}-z_{a},  \label{yasno} \\
&&  \notag \\
Q &=&ac^{2}+b^{2}c+a^{2}z_{aa}+2abz_{ab}+b^{2}z_{bb}-2(az_{a}+bz_{b}-z). 
\notag
\end{eqnarray}

\textbf{2}. Let us consider the Egorov hydrodynamic type system (\textbf{\ref%
{23}}) with the local Hamiltonian structure 
\begin{equation}
a_{t}=\partial _{x}\frac{\partial h_{1}}{\partial a}\text{, \ \ \ }%
b_{t}=\partial _{x}\frac{\partial h_{1}}{\partial c}\text{, \ \ \ }%
c_{t}=\partial _{x}\frac{\partial h_{1}}{\partial b}\text{,}  \label{26}
\end{equation}%
where $\partial h_{1}/\partial a=b$ (see (\textbf{\ref{1}})). Then we are
able to choose the extra commuting flow (\textbf{\ref{25}}) written in the
same Hamiltonian form%
\begin{equation*}
a_{y}=\partial _{x}\frac{\partial h_{2}}{\partial a}\text{, \ \ \ }%
b_{y}=\partial _{x}\frac{\partial h_{2}}{\partial c}\text{, \ \ \ }%
c_{y}=\partial _{x}\frac{\partial h_{2}}{\partial b}\text{,}
\end{equation*}%
where $\partial h_{2}/\partial a=c$ (see (\textbf{\ref{eg}})). Since $%
\partial h_{2}/\partial c=\partial h_{1}/\partial b$, both Egorov
hydrodynamic type systems are determined by the sole function $z(b,c)$
satisfying the famous \textit{associativity equation }(see \textbf{\cite%
{Dubr}}, \textbf{\cite{Fer+centre}}, \textbf{\cite{Yavuz}})%
\begin{equation*}
1+z_{bbc}z_{bcc}=z_{bbb}z_{ccc},
\end{equation*}%
where $h_{1}=ab+z_{c}$, $h_{2}=ac+z_{b}$ and the momentum density $P$ is
given by $h_{0}=a^{2}/2+bc$.

\textbf{Theorem}: \textit{The hydrodynamic type system} (\textbf{\ref{ega}}) 
\textit{is equivalent to the hydrodynamic type system} (\textbf{\ref{26}}) 
\textit{under the transformation} $x\leftrightarrow t$.

\textbf{Proof}: Let us replace $x\leftrightarrow t$, $\partial
h_{1}/\partial c\rightarrow \tilde{a}$, $\partial h_{1}/\partial
b\rightarrow \tilde{c}$, $\partial h_{1}/\partial a\rightarrow \tilde{b}$; $%
c\rightarrow \partial \tilde{h}_{1}/\partial \tilde{a}$, $a\rightarrow
\partial \tilde{h}_{1}/\partial \tilde{b}$, $b\rightarrow \partial \tilde{h}%
_{1}/\partial \tilde{c}$. Then (\textbf{\ref{26}}) transforms in (\textbf{%
\ref{ega}}).

\textbf{Remark}: The transformation connecting the above associativity
equations was found in \textbf{\cite{Yavuz}}).

\textbf{3}. In general case the Egorov hydrodynamic type system (\textbf{\ref%
{ham}}) has the couple of extra conservation laws (\textbf{\ref{1}}), where $%
a$ and $b$ are not connected with the Hamiltonian structure (\textbf{\ref%
{ham}}).

Let us consider the commuting flow (see (\textbf{\ref{ham}}) and (\textbf{%
\ref{zima}}))%
\begin{equation*}
\partial _{z}a_{i}=\partial _{x}(\partial \tilde{h}/\partial a^{i}),
\end{equation*}%
where the extra conservation law is (see (\textbf{\ref{1}}))%
\begin{equation}
\partial _{z}a=\partial _{x}a_{1}.  \label{ext}
\end{equation}

Under the transformation (see \textbf{\cite{Maks+notice}}) $x\leftrightarrow
z$, the above commuting flow reduces to the Egorov hydrodynamic type system%
\begin{equation}
\partial _{x}c_{i}=\partial _{z}(\partial \bar{h}/\partial c^{i}),
\label{exp}
\end{equation}%
where $c_{i}=\partial \tilde{h}/\partial a^{k}$, $\bar{h}=a^{k}\partial 
\tilde{h}/\partial a^{k}-\tilde{h}$ and the extra conservation law is (cf. (%
\textbf{\ref{ext}}))%
\begin{equation*}
\partial _{x}(\partial \bar{h}/\partial c^{1})=\partial _{z}a,
\end{equation*}%
where $a_{1}=\partial \bar{h}/\partial c^{1}$. Thus, in this case the
potential of the Egorov metric $\bar{a}\equiv c_{1}$. At the same time the
Egorov hydrodynamic type system (\textbf{\ref{exp}}) possesses the commuting
flow%
\begin{equation*}
\partial _{y}c_{i}=\partial _{z}(\partial \text{\b{h}/}\partial c^{i}),
\end{equation*}%
where the extra conservation law is%
\begin{equation*}
\partial _{y}c_{1}=\partial _{z}c_{2}.
\end{equation*}%
Thus, this general case is reduced to the simplest case described above.

\section{Reciprocal Transformations}

In the previous section we considered the simplest reciprocal transformation%
\begin{equation*}
dy^{k}=\sigma _{n}^{k}dt^{n}\text{, \ \ \ \ \ \ }\sigma _{n}^{k}=\limfunc{%
const}
\end{equation*}%
preserving the Egorov hydrodynamic type systems. In this section a more
complicated reciprocal transformation is presented. Suppose we already know
(see, for instance, the previous section) all conservation laws and
commuting flows for the Egorov hydrodynamic type systems (\textbf{\ref{sima}}%
). The generalized reciprocal transformation (see \textbf{\cite{FerMax}})
contains $M$ rows and $M$ columns determined by $M$ conservation laws and $%
M-1$ commuting flows. The number $M$ and the number $N$ (see (\textbf{\ref{2}%
})) do not correlate to each other in general case. The Egorov hydrodynamic
type system (\textbf{\ref{sima}}) can be written in the potential symmetric
form (\textbf{\ref{18}}). If $M\geqslant N$, then the generalized symmetric
reciprocal transformation (cf. (\textbf{\ref{18}}))%
\begin{equation*}
d\tilde{y}_{i}=a_{ik}(\mathbf{a})dt^{k}
\end{equation*}%
preserves a potential symmetric form and again transforms (\textbf{\ref{sima}%
}) to the Egorov hydrodynamic type systems. If $M<N$, then the generalized
reciprocal transformation%
\begin{equation*}
d\tilde{y}_{i}=a_{ik}(\mathbf{a})dt^{k}\text{, \ \ \ \ \ \ }i=1,2,...,M\text{%
; \ \ \ \ \ \ \ }d\tilde{y}_{i}=\sigma _{ik}dt^{k}\text{, \ \ \ \ \ }%
i=M+1,...,N
\end{equation*}%
must contain a symmetric part including the potential $a$ of the Egorov
metric, while all other independent variables transform linearly (i.e. $%
\sigma _{ik}=\limfunc{const}$).

Without lost of generality we restrict our consideration for simplicity on
the reciprocal transformation%
\begin{equation}
dy^{1}=a_{k}dt^{k}\text{, \ \ \ \ \ \ \ \ }dy^{k}=dt^{k}\text{, \ \ \ \ }%
k=2,3,...,N.  \label{gen}
\end{equation}%
Then the Egorov hydrodynamic type system (\textbf{\ref{sima}}) reduces to
the similar set of the Egorov hydrodynamic type systems%
\begin{equation*}
\partial _{y^{k}}\left( -\frac{1}{a_{1}}\right) =\partial _{y^{1}}\frac{a_{k}%
}{a_{1}}\text{, \ \ \ \ \ \ \ \ }\partial _{y^{k}}\frac{a_{n}}{a_{1}}%
=\partial _{y^{1}}\left( a_{nk}-\frac{a_{n}a_{k}}{a_{1}}\right) \text{, \ \
\ }k,n=2,3,...,N.
\end{equation*}%
A recalculation of local Hamiltonian structures under generalized reciprocal
transformation is given in \textbf{\cite{FerMax}}. In this section we
restrict our consideration on the above three component case (\textbf{\ref%
{ega}}).

\textbf{Theorem}: \textit{The local Hamiltonian structure }(\textbf{\ref{ega}%
}) \textit{of the Egorov hydrodynamic type system} (\textbf{\ref{23}}) 
\textit{is invariant under the reciprocal transformation} (\textbf{\ref{gen}}%
)%
\begin{equation}
dz=adx+bdt\text{, \ \ \ \ \ \ \ \ }dy=dt\text{.}  \label{rt}
\end{equation}

\textbf{Proof}: The Egorov hydrodynamic type system (\textbf{\ref{23}}) has $%
4$ local conservation laws associated with local Hamiltonian structure (%
\textbf{\ref{ega}}), where the momentum density $P$ is the quadratic
expression with respect to flat coordinates (see \textbf{\cite{Maks+Tsar}})%
\begin{equation}
P=ac+\frac{b^{2}}{2}.  \label{imp}
\end{equation}%
Under the above reciprocal transformation \textit{any} conservation law
density $h$ reduces to $h/a^{1}$. Thus, (\textbf{\ref{imp}}) reduces to%
\begin{equation*}
\tilde{P}=\tilde{a}\tilde{c}+\frac{\tilde{b}^{2}}{2},
\end{equation*}%
where $\tilde{P}=-c/a$, $\tilde{a}=1/a$, $\tilde{c}=-P/a$, $\tilde{b}=-b/a$.

\textbf{Remark}: The Hamiltonian density of the Egorov hydrodynamic type
system (\textbf{\ref{ega}})%
\begin{equation*}
\tilde{a}_{t}=\partial _{z}\frac{\partial \tilde{h}}{\partial \tilde{c}}%
\text{, \ \ \ \ \ \ \ }\tilde{b}_{t}=\partial _{z}\frac{\partial \tilde{h}}{%
\partial \tilde{b}}\text{, \ \ \ \ \ \ \ }\tilde{c}_{t}=\partial _{z}\frac{%
\partial \tilde{h}}{\partial \tilde{a}}
\end{equation*}%
is given by $\tilde{h}=h/a$.

\section{Three orthogonal Egorov curvilinear coordinate systems}

$N$ orthogonal Egorov curvilinear coordinate nets are described by the
Bianchi--Darboux--Egorov--Lame system (\textbf{\ref{9}})%
\begin{equation*}
\partial _{i}\beta _{jk}=\beta _{ji}\beta _{ik}\text{, \ \ \ \ }i\neq j\neq k%
\text{; \ \ \ \ \ \ \ }\beta _{ik}=\beta _{ki}\text{, \ \ \ \ }i\neq k\text{%
, \ \ \ \ \ \ \ }\delta \beta _{ik}=0.
\end{equation*}%
$N$ orthogonal curvilinear coordinate nets were investigated in many
publications (see, for instance, \textbf{\cite{Bianchi}}, \textbf{\cite%
{Darboux}}, \textbf{\cite{Ganzha}}, \textbf{\cite{Tsar}}, \textbf{\cite%
{Zakh+ort}}; and plenty references therein). In this section we establish a
new link connecting an infinite set of $N$ orthogonal Egorov curvilinear
coordinate systems:%
\begin{equation}
...\leftarrow \beta _{ik}^{(-2)}\leftarrow \beta _{ik}^{(-1)}\leftarrow
\beta _{ik}^{(0)}\rightarrow \beta _{ik}^{(1)}\rightarrow \beta
_{ik}^{(2)}\rightarrow ...  \label{ort}
\end{equation}%
Without lost of generality we restrict our consideration for simplicity on a
three component case only. The main advantage of the approach presented
below is that all formulas are given via \textbf{flat} coordinates only (in
comparison with the Riemann invariants). In such a case corresponding
formulas can be easily used for a construction of solutions for the WDVV\
equation.

Suppose all rotations coefficients $\beta _{ik}^{(0)}$ and the Lame
coefficients $H_{(k)i}$ are given. It means that the couple of the Egorov
hydrodynamic type systems (\textbf{\ref{eg}}) is given too. The local
Hamiltonian structure (\textbf{\ref{ham}}) can be reduced (in three
component case) to (\textbf{\ref{ega}}) by a linear transformation of field
variables ($a^{k}\rightarrow \sigma _{s}^{k}a^{s}$, $\sigma _{s}^{k}=%
\limfunc{const}$), because the symmetric constant \textit{non-degenerate }%
metric $\bar{g}^{ik}$ is reducible to the diagonal or skew-diagonal form.

\textbf{1}. Let us apply the reciprocal transformation (\textbf{\ref{gen}})%
\begin{eqnarray}
dx^{(1)} &=&a^{(0)}dx^{(0)}+b^{(0)}dt^{(0)}+c^{(0)}dy^{(0)}+\left(
a^{(0)}c^{(0)}+\frac{\left( b^{(0)}\right) ^{2}}{2}\right) d\tau ^{(0)}\text{%
,}  \notag \\
&&  \label{e} \\
dt^{(1)} &=&dt^{(0)}\text{, \ \ \ \ \ \ \ \ \ \ \ }d\tau ^{(1)}=dy^{(0)}%
\text{,\ \ \ \ \ \ \ \ \ \ \ \ }dy^{(1)}=d\tau ^{(0)}  \notag
\end{eqnarray}%
to the Egorov hydrodynamic type systems (\textbf{\ref{eg}}) (see also (%
\textbf{\ref{ega}}), (\textbf{\ref{ego}})) written for simplicity in the
potential symmetric form (\textbf{\ref{tima}})%
\begin{equation*}
d\left( 
\begin{array}{c}
x^{(1)} \\ 
\tilde{t}^{(1)} \\ 
\tilde{y}^{(1)} \\ 
\tilde{\tau}^{(1)}%
\end{array}%
\right) =\left( 
\begin{array}{cccc}
a^{(0)} & b^{(0)} & c^{(0)} & P^{(0)} \\ 
b^{(0)} & u^{(0)} & v^{(0)} & R^{(0)} \\ 
c^{(0)} & v^{(0)} & w^{(0)} & S^{(0)} \\ 
P^{(0)} & R^{(0)} & S^{(0)} & Q^{(0)}%
\end{array}%
\right) d\left( 
\begin{array}{c}
x^{(0)} \\ 
t^{(0)} \\ 
y^{(0)} \\ 
\tau ^{(0)}%
\end{array}%
\right) ,
\end{equation*}%
where $z^{(0)}$ satisfies the associativity equation (\textbf{\ref{ace}}).
The hydrodynamic type systems (\textbf{\ref{eg}}) form the canonical Egorov
basic set (\textbf{\ref{ega}}), (\textbf{\ref{ego}}) (i.e. the fluxes $%
b^{(0)}$, $c^{(0)}$ of the first conservation law $\partial
_{t^{(0)}}a^{(0)}=\partial _{x^{(0)}}b^{(0)}$, $\partial
_{y^{(0)}}a^{(0)}=\partial _{x^{(0)}}c^{(0)}$ are the corresponding flat
coordinates). Since $a^{(1)}=1/a^{(0)}$, $b^{(1)}=-b^{(0)}/a^{(0)}$, $%
c^{(1)}=-P^{(0)}/a^{(0)}$ are flat coordinates (and $a^{(1)}$ is a potential
of the \textit{transformed} Egorov metric), then the \textit{transformed}
canonical Egorov basic set (\textbf{\ref{sima}}) can be extracted from the
above \textit{transformed} potential symmetric form%
\begin{equation*}
d\left( 
\begin{array}{c}
x^{(0)} \\ 
-\tilde{t}^{(1)} \\ 
-\tilde{\tau}^{(1)} \\ 
-\tilde{y}^{(1)}%
\end{array}%
\right) =\left( 
\begin{array}{cccc}
1/a^{(0)} & -\frac{b^{(0)}}{a^{(0)}} & -\frac{P^{(0)}}{a^{(0)}} & -\frac{%
c^{(0)}}{a^{(0)}} \\ 
-\frac{b^{(0)}}{a^{(0)}} & \frac{\left( b^{(0)}\right) ^{2}}{a^{(0)}}-u^{(0)}
& \frac{b^{(0)}P^{(0)}}{a^{(0)}}-R^{(0)} & \frac{b^{(0)}c^{(0)}}{a^{(0)}}%
-v^{(0)} \\ 
-\frac{P^{(0)}}{a^{(0)}} & \frac{b^{(0)}P^{(0)}}{a^{(0)}}-R^{(0)} & \frac{%
\left( P^{(0)}\right) ^{2}}{a^{(0)}}-Q^{(0)} & \frac{c^{(0)}P^{(0)}}{a^{(0)}}%
-S^{(0)} \\ 
-\frac{c^{(0)}}{a^{(0)}} & \frac{b^{(0)}c^{(0)}}{a^{(0)}}-v^{(0)} & \frac{%
c^{(0)}P^{(0)}}{a^{(0)}}-S^{(0)} & \frac{\left( c^{(0)}\right) ^{2}}{a^{(0)}}%
-w^{(0)}%
\end{array}%
\right) d\left( 
\begin{array}{c}
x^{(1)} \\ 
t^{(1)} \\ 
\tau ^{(1)} \\ 
y^{(1)}%
\end{array}%
\right) .
\end{equation*}%
The transformed canonical Egorov basic set must be given by (see (\textbf{%
\ref{ega}}), (\textbf{\ref{ego}}))%
\begin{equation}
\begin{array}{ccccccccc}
\partial _{t^{(1)}}a^{(1)}=\partial _{x^{(1)}}b^{(1)}, &  &  &  &  &  &  & 
& \partial _{y^{(1)}}a^{(1)}=\partial _{x^{(1)}}c^{(1)}, \\ 
&  &  &  &  &  &  &  &  \\ 
\partial _{t^{(1)}}b^{(1)}=\partial _{x^{(1)}}\left(
c^{(1)}+z_{b^{(1)}b^{(1)}}^{(1)}\right) , &  &  &  &  &  &  &  & \partial
_{y^{(1)}}b^{(1)}=\partial _{x^{(1)}}z_{a^{(1)}b^{(1)}}^{(1)}, \\ 
&  &  &  &  &  &  &  &  \\ 
\partial _{t^{(1)}}c^{(1)}=\partial _{x^{(1)}}z_{a^{(1)}b^{(1)}}^{(1)}, &  & 
&  &  &  &  &  & \partial _{y^{(1)}}c^{(1)}=\partial
_{x^{(1)}}z_{a^{(1)}a^{(1)}}^{(1)},%
\end{array}
\label{n}
\end{equation}%
where $z^{(1)}$ satisfies the associativity equation (\textbf{\ref{ace}}).
Comparing corresponding fluxes from the above equations%
\begin{equation*}
z_{a^{(1)}a^{(1)}}^{(1)}=\frac{(P^{(0)})^{2}}{a^{(0)}}-Q^{(0)}\text{, \ \ \
\ \ \ \ }z_{a^{(1)}b^{(1)}}^{(1)}=\frac{b^{(0)}P^{(0)}}{a^{(0)}}-R^{(0)}%
\text{, \ \ \ \ \ \ \ }c^{(1)}+z_{b^{(1)}b^{(1)}}^{(1)}=\frac{(b^{(0)})^{2}}{%
a^{(0)}}-u^{(0)},
\end{equation*}%
one can compute the new solution $z^{(1)}$ of the associativity equation (%
\textbf{\ref{ace}}) in quadratures%
\begin{equation*}
dz^{(1)}=z_{a^{(1)}}^{(1)}d\frac{1}{a^{(0)}}-z_{b^{(1)}}^{(1)}d\frac{b^{(0)}%
}{a^{(0)}},
\end{equation*}%
where%
\begin{eqnarray*}
dz_{a^{(1)}}^{(1)} &=&\left( \frac{\left( P^{(0)}\right) ^{2}}{a^{(0)}}%
-Q^{(0)}\right) d\frac{1}{a^{(0)}}-\left( \frac{b^{(0)}P^{(0)}}{a^{(0)}}%
-R^{(0)}\right) d\frac{b^{(0)}}{a^{(0)}}, \\
&& \\
dz_{b^{(1)}}^{(1)} &=&\left( \frac{b^{(0)}P^{(0)}}{a^{(0)}}-R^{(0)}\right) d%
\frac{1}{a^{(0)}}-\left( \frac{(b^{(0)})^{2}}{a^{(0)}}-u^{(0)}+\frac{P^{(0)}%
}{a^{(0)}}\right) d\frac{b^{(0)}}{a^{(0)}}.
\end{eqnarray*}%
Moreover, the substitution (\textbf{\ref{yasno}}) in the above differentials
yields an explicit link%
\begin{equation}
z^{(1)}=\frac{(b^{(0)})^{4}}{8(a^{(0)})^{3}}-\frac{1}{(a^{(0)})^{2}}z^{(0)}.
\label{link}
\end{equation}%
of the solutions $z^{(0)}$ and $z^{(1)}$ of the associativity equation (%
\textbf{\ref{ace}}).

\textbf{Remark}: \textit{The Ribaucour transformation}. Transformations of
local Hamiltonian structures under generalized reciprocal transformations
are described in \textbf{\cite{FerMax}}. In the simplest case (\textbf{\ref%
{e}}) the comparison of the Lame coefficients $\bar{H}_{(1)i}^{(0)}$ and the 
\textit{transformed} Lame coefficients $\bar{H}_{(1)i}^{(1)}$ from (see (%
\textbf{\ref{19}}))%
\begin{equation*}
\partial _{i}a^{(0)}=(\bar{H}_{(1)i}^{(0)})^{2}\text{, \ \ \ \ \ \ \ }%
\partial _{i}a^{(1)}\equiv \partial _{i}\frac{1}{a^{(0)}}=-\frac{1}{%
(a^{(0)})^{2}}(\bar{H}_{(1)i}^{(0)})^{2}=(\bar{H}_{(1)i}^{(1)})^{2}
\end{equation*}%
leads to the Ribaucour transformation (see details in \textbf{\cite{Tsar}})
of the three orthogonal curvilinear coordinate nets%
\begin{equation*}
\beta _{ik}^{(1)}=\beta _{ik}^{(0)}-\frac{\bar{H}_{(1)i}^{(0)}\bar{H}%
_{(1)k}^{(0)}}{a^{(0)}}\text{, \ \ \ \ \ \ \ \ \ \ \ }\bar{H}_{(1)k}^{(1)}=i%
\frac{\bar{H}_{(1)k}^{(0)}}{a^{(0)}}.
\end{equation*}%
These rotation coefficients $\beta _{ik}^{(0)}$ and $\beta _{ik}^{(0)}$ are
symmetric and satisfy the zero curvature condition (\textbf{\ref{17}}).
Thus, indeed, the Egorov three orthogonal curvilinear coordinate nets are
preserved under the above Ribaucour transformation (see \textbf{\cite%
{Bianchi}} and \textbf{\cite{Tsar}}).

\textbf{2}. Let us rewrite the canonical Egorov basic set (\textbf{\ref{n}})
in the form%
\begin{equation*}
\begin{array}{ccccccccc}
\partial _{t^{(1)}}z_{a^{(1)}a^{(1)}}^{(1)}=\partial
_{y^{(1)}}z_{a^{(1)}b^{(1)}}^{(1)}, &  &  &  &  &  &  &  & \partial
_{x^{(1)}}z_{a^{(1)}a^{(1)}}^{(1)}=\partial _{y^{(1)}}c^{(1)}, \\ 
&  &  &  &  &  &  &  &  \\ 
\partial _{t^{(1)}}z_{a^{(1)}b^{(1)}}^{(1)}=\partial _{y^{(1)}}\left(
c^{(1)}+z_{b^{(1)}b^{(1)}}^{(1)}\right) , &  &  &  &  &  &  &  & \partial
_{x^{(1)}}z_{a^{(1)}b^{(1)}}^{(1)}=\partial _{y^{(1)}}b^{(1)}, \\ 
&  &  &  &  &  &  &  &  \\ 
\partial _{t^{(1)}}c^{(1)}=\partial _{y^{(1)}}b^{(1)}, &  &  &  &  &  &  & 
& \partial _{x^{(1)}}c^{(1)}=\partial _{y^{(1)}}a^{(1)}.%
\end{array}%
\end{equation*}%
Thus, under the linear transformation of independent variables%
\begin{equation}
x^{(2)}=y^{(1)}\text{, \ \ \ \ \ }t^{(2)}=t^{(1)}\text{, \ \ \ \ \ \ }%
y^{(2)}=x^{(1)}  \label{lina}
\end{equation}%
the above canonical Egorov basic set reduces to the \textit{transformed}
canonical Egorov basic set%
\begin{equation*}
\begin{array}{ccccccccc}
\partial _{t^{(2)}}a^{(2)}=\partial _{x^{(2)}}b^{(2)}, &  &  &  &  &  &  & 
& \partial _{y^{(2)}}a^{(2)}=\partial _{x^{(2)}}c^{(2)}, \\ 
&  &  &  &  &  &  &  &  \\ 
\partial _{t^{(2)}}b^{(2)}=\partial _{x^{(2)}}\left(
c^{(2)}+z_{b^{(2)}b^{(2)}}^{(2)}\right) , &  &  &  &  &  &  &  & \partial
_{y^{(2)}}b^{(2)}=\partial _{x^{(2)}}z_{a^{(2)}b^{(2)}}^{(2)}, \\ 
&  &  &  &  &  &  &  &  \\ 
\partial _{t^{(2)}}c^{(2)}=\partial _{x^{(2)}}z_{a^{(2)}b^{(2)}}^{(2)}, &  & 
&  &  &  &  &  & \partial _{y^{(2)}}c^{(2)}=\partial
_{x^{(2)}}z_{a^{(2)}a^{(2)}}^{(2)},%
\end{array}%
\end{equation*}%
where%
\begin{equation*}
a^{(2)}=z_{a^{(1)}a^{(1)}}^{(1)}\text{, \ \ }b^{(2)}=z_{a^{(1)}b^{(1)}}^{(1)}%
\text{, \ \ }c^{(2)}=c^{(1)}\text{, \ \ }%
z_{b^{(2)}b^{(2)}}^{(2)}=z_{b^{(1)}b^{(1)}}^{(1)}\text{, \ \ }%
z_{a^{(2)}b^{(2)}}^{(2)}=b^{(1)}\text{, \ \ }%
z_{a^{(2)}a^{(2)}}^{(2)}=a^{(1)}.
\end{equation*}%
It means that the solution $z^{(2)}$ of the associativity equation (\textbf{%
\ref{ace}}) can be found in quadratures%
\begin{equation}
dz^{(2)}=z_{a^{(2)}}^{(2)}dz_{a^{(1)}a^{(1)}}^{(1)}+z_{b^{(2)}}^{(2)}dz_{a^{(1)}b^{(1)}}^{(1)},
\label{shik}
\end{equation}%
where%
\begin{equation*}
z_{a^{(2)}}^{(2)}=a^{(1)}z_{a^{(1)}a^{(1)}}^{(1)}+b^{(1)}z_{a^{(1)}b^{(1)}}^{(1)}-z_{a^{(1)}}^{(1)}%
\text{, \ \ \ \ \ \ \ \ }%
dz_{b^{(2)}}^{(2)}=z_{b^{(1)}b^{(1)}}^{(1)}dz_{a^{(1)}b^{(1)}}^{(1)}+b^{(1)}dz_{a^{(1)}a^{(1)}}^{(1)}.
\end{equation*}

\textbf{Remark}: The above right differential%
\begin{equation*}
dG=z_{bb}dz_{ab}+bdz_{aa}
\end{equation*}%
exists iff the function $z$ is a solution of the associativity equation (%
\textbf{\ref{ace}}). This function $G$ is the first such example of
infinitely many expressions, which cannot be computed explicitly. A
computation of higher conservation laws leads to similar differentials. For
instance (see (\textbf{\ref{ega}}), (\textbf{\ref{ego}})),%
\begin{equation*}
\partial _{y}(bc+z_{b})=\partial _{x}(cz_{ab}+G)\text{, \ \ \ \ \ \ }%
\partial _{t}\left( \frac{c^{2}}{2}+z_{a}\right) =\partial
_{x}[z_{ab}(c+z_{bb})+bz_{aa}-G].
\end{equation*}

Thus, the \textit{iterative replication} of solutions for the associativity
equation (\textbf{\ref{ace}}) can be based on the above two steps%
\begin{equation}
...\leftarrow z^{(-2)}\leftarrow z^{(-1)}\leftarrow z^{(0)}\rightarrow
z^{(1)}\rightarrow z^{(2)}\rightarrow ...  \label{tort}
\end{equation}%
The \textit{negative} direction means that the transformations (\textbf{\ref%
{e}}) and (\textbf{\ref{lina}}) are applied in an inverse order.

The \textbf{main statement} of this section: \textit{Each solution of the
associativity equation} (\textbf{\ref{ace}}) \textit{creates the canonical
Egorov basic set, simultaneously, the corresponding basic Lame coefficients} 
$\bar{H}_{i}^{(k)}$ \textit{and rotation coefficients} $\beta _{ik}$. 
\textit{It means, that an infinite set of solutions} $z^{(k)}$ (\textit{see}
(\textbf{\ref{link}}), (\textbf{\ref{shik}}) \textit{and} (\textbf{\ref{tort}%
})) \textit{creates an infinite set of orthogonal curvilinear coordinate nets%
} (\textbf{\ref{ort}}). \textit{Thus, a description of all solutions of the
associativity equation} (\textbf{\ref{ace}}) \textit{is equivalent to a
description of three orthogonal curvilinear coordinate nets}.

This set of transformations has the same interpretation as the so-called
``dressing method'' in the soliton theory.

Choosing the initial solution of the associativity equation (\textbf{\ref%
{ace}}) $z^{(0)}=0$, the first iteration yields the solution given by%
\begin{equation*}
z^{(1)}=\frac{b^{4}}{8a}.
\end{equation*}%
The second iteration yields the same solution. The fourth iteration yields
again the ``zero'' solution. However, in general case this iterative chain
of transformations cannot be truncated.

In this paper just a three component case was considered in details, but
without any restrictions this approach can be applied for $N$ component case.

\textbf{Remark}: The first transformation (\textbf{\ref{link}}) is exactly
the ``inversion $I$'' transformation for the WDVV equation found in \textbf{%
\cite{Dubr}} (see the formulas B.13, Appendix \textbf{B}). The second
transformation is exactly the Legendre type transformation for the WDVV
equation found in \textbf{\cite{Dubr}} (see the formulas B.1 and B.2,
Appendix \textbf{B}).

\section{Nonlocal Hamiltonian structures}

Theory of nonlocal Hamiltonian structures for hydrodynamic type systems was
constructed by E.V. Ferapontov in \textbf{\cite{Fer+trans}} (see also 
\textbf{\cite{Malt+Nov}}). The simplest nonlocal Hamiltonian structure is
associated with the metric of constant curvature (see also \textbf{\cite%
{Fer+Mokh}}, \textbf{\cite{Maks+cc}}). The existence of the Hamiltonian
structure for the Egorov hydrodynamic type systems (\textbf{\ref{sima}})
leads to the symmetry operator (see (\textbf{\ref{symmetry}}) for the local
Hamiltonian structure). In this section without lost of generality we
restrict our consideration for simplicity on the nonlocal Hamiltonian
structure associated with the metric of constant curvature. The
corresponding hydrodynamic type system can be written via special field
variables $a^{k}$ in the conservative form (cf. (\textbf{\ref{ham}}))%
\begin{equation*}
a_{t}^{i}=\partial _{x}\left[ (\bar{g}^{ik}-\varepsilon a^{i}a^{k})\frac{%
\partial h}{\partial a^{k}}+\varepsilon a^{i}h\right] ,
\end{equation*}%
where $\bar{g}^{ik}$ is a constant symmetric matrix and $\varepsilon $ is a
constant curvature.

Let us apply the reciprocal transformation (\textbf{\ref{rt}}) to the Egorov
hydrodynamic type system (\textbf{\ref{26}}). Then the quadratic
relationship (connected with local Hamiltonian structure, see \textbf{\cite%
{Maks+Tsar}})%
\begin{equation*}
P=bc+\frac{a^{2}}{2}
\end{equation*}%
transforms in another type quadratic relationship (connected with the above
nonlocal Hamiltonian structure; see \textbf{\cite{Malt+Nov}} and \textbf{%
\cite{Maks+cc}})%
\begin{equation*}
\frac{1}{2}=a_{1}a_{2}-a_{3}a_{4},
\end{equation*}%
where $a_{1}=1/a$, $a_{2}=P/a$, $a_{3}=-b/a$, $a_{4}=-c/a$. The coordinate $%
a_{1}$ is the potential of the Egorov metric, and the corresponding
transformed Egorov hydrodynamic type system has the above nonlocal
Hamiltonian structure. Thus, this nonlocal Hamiltonian structure is
reducible to the local Hamiltonian structure (\textbf{\ref{ham}}).

\section*{Acknowledgement}

I thank Eugeni Ferapontov, John Gibbons, Yuji Kodama and Sergey Tsarev for
their help and clarifying discussions.

I am grateful to the Institute of Mathematics in Taipei (Taiwan) where some
part of this work has been done, and especially to Jen-Hsu Chang, Jyh-Hao
Lee, Ming-Hien Tu and Derchyi Wu for fruitful discussions.%


\addcontentsline{toc}{section}{References}

\begin{thebibliography}{99}
\bibitem{Bianchi} \emph{L. Bianchi}, \newblock Sisteme tripli ortogonali,
Ed. Cremonese, Roma (1955).

\bibitem{Darboux} \emph{G. Darboux}, \newblock Lecons sur les systemes
orthogonaux et les coordonnes curvilignes. Paris, Gautier-Villar, 1910.

\bibitem{Dubr} \emph{B.A. Dubrovin,} \newblock Integrable systems in
topological field theory, Nucl. Phys. B, \textbf{379} (1992) 627-689. \emph{%
B.A. Dubrovin,} \newblock Hamiltonian formalism of Whitham-type hierarchies
and topological Landau-Ginsburg models, Comm. Math. Phys., \textbf{145}
(1992) 195-207. \emph{B.A. Dubrovin}, \newblock Geometry of 2D topological
field theories, Lecture Notes in Math. 1620, Springer-Verlag (1996) 120-348.

\bibitem{Dubr+Nov} \emph{B.~A. Dubrovin and S.~P. Novikov,} \newblock %
Hamiltonian formalism of one-dimensional systems of hydrodynamic type and
the Bogolyubov-Whitham averaging method, Soviet Math. Dokl., \textbf{27}
(1983) 665--669. \emph{B.~A. Dubrovin and S.~P. Novikov,} \newblock %
Hydrodynamics of weakly deformed soliton lattices. Differential geometry and
Hamiltonian theory, Russian Math. Surveys, 44:6 (1989) 35--124.

\bibitem{Egorov} \emph{D.F. Egorov}, \newblock Works in Differential
Geometry. Moscow, Nauka, 1970.

\bibitem{Fer+centre} \emph{E.~V. Ferapontov,} \newblock Hypersurfaces with
flat centroaffine metric and equations of associativity, Geometriae
Dedicata, \textbf{103} (2004) 33-49.

\bibitem{Fer+trans} \emph{E.~V. Ferapontov,} \newblock Nonlocal Hamiltonian
operators of hydrodynamic type: differential geometry and applications,
Amer. Math. Soc. Transl. (2), \textbf{170} (1995) 33-58.

\bibitem{Yavuz} \emph{E.V. Ferapontov, C.A.P. Galvao, O.I. Mokhov, Y. Nutku}%
, \newblock Bi-Hamiltonian structure of equations of associativity in 2-d
topological field theory, Comm. Math. Phys., \textbf{186} (1997) 649-669. 
\emph{E.V. Ferapontov, O.I. Mokhov}, \newblock Equations of associativity of
two-dimensional topological field theory as integrable Hamiltonian
nondiagonalisable systems of hydrodynamic type, Funct. Anal. and it's Appl., 
\textbf{30} No.3 (1996) 62-72.

\bibitem{FerKarMax} \emph{E.V. Ferapontov, K.R. Khusnutdinova, M.V. Pavlov},
Classification of integrable (2+1) dimensional quasilinear hierarchies.
Theor. Math. Phys. \textbf{144} No. 1 (2005) 35-43.

\bibitem{Fer+Mokh} \emph{E.V. Ferapontov, O.I. Mokhov}, \newblock Nonlocal
Hamiltonian operators of hydrodynamic type that are connected with metrics
of constant curvature, Russian Math. Surveys, \textbf{45} No. 3 (1990)
218--219.

\bibitem{FerMax} \emph{E.V. Ferapontov, M.V. Pavlov}, \newblock Reciprocal
transformations of Hamiltonian operators of hydrodynamic type: nonlocal
Hamiltonian formalism for linearly degenerate systems, J. Math. Phys., 
\textbf{44} No. 3 (2003) 1150-1172.

\bibitem{Ganzha} \emph{E.I. Ganzha, S.P. Tsarev}, \newblock An algebraic
formula for superposition and completeness of the Backlund transformations
of ($2+1$)-dimensional integrable systems, Russian Math. Surveys, \textbf{51}
6 (1996), 1200-1202.

\bibitem{Gibbons} \emph{J. Gibbons}, \newblock Collisionless Boltzmann
equations and integrable moment equations, Physica\textbf{\ D} \textbf{3},
No. 3 (1981), 503--511.

\bibitem{Gib+Kod} \emph{J. Gibbons and Y. Kodama}, \newblock A method for
solving the dispersionless KP hierarchy and its exact solutions, II. Phys.
Lett. A \textbf{135} No. 3 (1989), 167--170. \emph{Y. Kodama}, \newblock %
Solution of the dispersionless Toda equation, Phys. Lett. A, \textbf{147}
(1990) 477-480.

\bibitem{Krich} \emph{I.M. Krichever,} \newblock Whitham theory for
integrable systems and topological quantum field theories. New symmetry
principles in quantum field theory (Cargse, 1991), 309--327, NATO Adv. Sci.
Inst. Ser. B Phys., 295, Plenum, New York, 1992. \emph{I.M. Krichever,} %
\newblock The $\tau $-function of the universal Whitham hierarchy, matrix
models and topological field theories, Comm. Pure Appl. Math., \textbf{47}
(1994) 437-475. \emph{I.M. Krichever, }\newblock Algebraic-geometrical $N$
orthogonal curvilinear coordinate systems and solutions of the associativity
equations. Funct. Anal. Appl., \textbf{31} No. 1 (1997) 25--39. \emph{A.A.
Akhmetshin, I.M. Krichever, Y.S. Volvovski,} \newblock A generating formula
for the solutions of the associativity equations. Russian Math. Surveys, 
\textbf{54} No. 2 (1999) 427--429.

\bibitem{Malt+Nov} \emph{A.Ya. Maltsev, S.P. Novikov}, \newblock On the
local systems Hamiltonian in the weakly nonlocal Poisson brackets, Physica
D, \textbf{156} (2001) 53-80.

\bibitem{Manas} \emph{M. Manas, L.M. Alonso, E. Medina}, \newblock On the
Whitham hierarchy: dressing scheme, string equations and additional
symmetries, J. Phys. A: Math. Gen., \textbf{39 }(2006) 2349. \emph{M. Manas,
L.M. Alonso, E. Medina}, \newblock Dressing methods for geometric nets: I.
Conjugate nets, J. Phys. A: Math. Gen., \textbf{33} (2000) 2871--2894. \emph{%
M. Manas, L.M. Alonso, E. Medina}, \newblock Dressing methods for geometric
nets: II. Orthogonal and Egorov nets, J. Phys. A: Math. Gen., \textbf{33}
(2000) 7181--7206. \emph{Q.P. Liu, M. Manas}, \newblock Symmetric reduction
of the vectorial fundamental transformation: application to the
Darboux--Egorov equations, J. Phys. A: Math. Gen., \textbf{32 }(1999)
5921--5927.

\bibitem{Nutku} \emph{Y. Nutku}, \newblock On a new class of completely
integrable nonlinear wave equations. Multi-Hamiltonian structure II, J.
Math. Phys., \textbf{28} No. 11 (1987) 2579--2585.

\bibitem{Maks+Benney} \emph{M.V. Pavlov, S.P. Tsarev}, \newblock %
Conservation laws for the Benney equations, Russian Math. Surveys, \textbf{46%
} No. 4 (1991) 196-197. \emph{M.V. Pavlov}, \newblock Local Hamiltonian
structures of Benney's system, Russian Phys. Dokl., \textbf{39} No. 9 (1994)
607-608. \emph{M.V. Pavlov}, \newblock Exact integrability of a system of
the Benney equations, Russian Phys. Dokl., \textbf{39} No. 11 (1994) 745-747.

\bibitem{Maks+kdv} \emph{M.V. Pavlov}, \newblock Whitham's averaging method
and the Korteweg-de Vries hierarchy, Russian Phys. Dokl., \textbf{39} No. 9
(1994) 615-617. \emph{M.V. Pavlov}, \newblock Hamiltonian structure of the
Whitham equations, Russian Math. Surveys, \textbf{49} No. 1 (1994) 241-242. 
\emph{M.V. Pavlov}, \newblock Multi-Hamiltonian structures of the Whitham
equations, Russian Acad. Sci. Dokl. Math., \textbf{50} No. 2 (1995) 220-223. 
\emph{M.V. Pavlov}, \newblock Dual Lagrangian representation of the KdV
equation and the general solution of the Whitham equations, Russian Acad.
Sci. Dokl. Math., \textbf{50} No. 3 (1995) 400-406.

\bibitem{Maks+notice} \emph{M.V. Pavlov}, \newblock Preservation of the
``form'' of Hamiltonian structures under linear changes of the independent
variables, Math. Notes, \textbf{57} No. 5-6 (1995) 489-495. \emph{S.P. Tsarev%
}, \newblock The Hamiltonian property of stationary and inverse equations of
condensed matter mechanics and mathematical physics, Math. Notes, \textbf{46}
No.1-2 (1989) 569-573.

\bibitem{Maks+cc} \emph{M.V. Pavlov}, \newblock Integrable systems and
metrics of constant curvature, Journal of Nonlinear Mathematical Physics.
No. 9 (2002) Supplement 1, 173-191.

\bibitem{Maks+Tsar} \emph{M. V. Pavlov, S.P. Tsarev,} \newblock
Three-Hamiltonian structures of the Egorov hydrodynamic type systems, Funct.
Anal. Appl., \textbf{37} No. 1 (2003) 32-45.

\bibitem{Yanenko} \emph{B.L. Rozhdestvenski, N.N. Yanenko}, \newblock %
Systems of quasilinear equations and their applications to gas dynamics.
Translated from the second Russian edition by J. R. Schulenberger.
Translations of Mathematical Monographs, 55. American Mathematical Society,
Providence, RI, 1983; Russian ed. Nauka, (1968) Moscow.

\bibitem{Tsar} \emph{S.P. Tsarev}, \newblock On Poisson brackets and
one-dimensional Hamiltonian systems of hydrodynamic type, Soviet Math.
Dokl., \textbf{31} (1985) 488--491. \emph{S.P. Tsarev}, \newblock The
geometry of Hamiltonian systems of hydrodynamic type. The generalized
hodograph method, Math. USSR Izvestiya, \textbf{37} No. 2 (1991) 397--419. 
\emph{S.P. Tsarev}, \newblock Classical differential geometry and
integrability of systems of hydrodynamic type. In: ``Applications of
Analytic and Geometrical Methods to Nonlinear Differential Equations'' ed.
P.A. Clarkson (Dordrecht: Kluwer) 1993. \emph{S.P. Tsarev}, \newblock %
Integrability of equations of hydrodynamic type from the end of the 19th to
the end of the 20th century. In: ``Integrability: the Seiberg-Witten and
Whitham equations'' (Edinburgh, 1998) p. 251--265, Gordon and Breach,
Amsterdam, 2000.

\bibitem{Zakh} \emph{V.E. Zakharov}, \newblock Benney's equations and
quasi-classical approximation in the inverse problem method, Funct. Anal.
Appl., \textbf{14} No. 2 (1980) 89-98. \emph{V.E. Zakharov}, \newblock On
the Benney's Equations, Physica 3D (1981) 193-200.

\bibitem{Zakh+ort} \emph{V.E. Zakharov}, \newblock Description of the $N$
orthogonal curvilinear coordinate systems and Hamiltonian integrable systems
of hydrodynamic type, I: Integration of the Lame equations. Duke Math. J. 
\textbf{94} No. 1 (1998) 103-139. \emph{V.E. Zakharov}, \newblock %
Integration of the Gauss-Codazzi equations, Teor. Math. Phys., \textbf{128}
No. 1 (2001) 946-956. \emph{V.E. Zakharov}, \newblock Application of the
Inverse Scattering Transform to Classical Problems of Differential Geometry
and General Relativity, Contemporary Mathematics, \textbf{301} (2002).
\end{thebibliography}
\end{document}